\documentclass[aps,prb,preprint]{revtex4}
\usepackage{graphicx}
\usepackage{amsmath}
\usepackage{amsfonts}
\usepackage{color}

\begin{document}

\title{Finite size and volume effects in line node semimetals: a first-principles investigation}

\author{Awadhesh Narayan}
\affiliation{Materials Theory, ETH Zurich, Wolfgang-Pauli-Strasse 27, CH 8093 Zurich, Switzerland.}

\date{\today}

\begin{abstract}
We systematically study the bulk and finite size effects on the band structure of prototypical line node materials using density functional theory based computations. For the bulk system, we analyze quantum oscillations with changes in the Fermi surface topology. We show that ultra thin slabs are gapped, with the first signatures of the line node appearing beyond a critical thickness. Further, motivated by possibility of tuning the bulk line node, we find that the line node radius is rather sensitive to bulk volume change. Based on this observation, we propose that application of pressure or suitable substrate engineering can be used as an effective means to control and tune the line node and the accompanying surface drumhead states.
\end{abstract}

\maketitle

\section{Introduction}
The study of topological materials, which originated with investigations of bulk insulators~\cite{hasan2010colloquium,qi2011topological}, has expanded to encompass gapless semimetals~\cite{bansil2016colloquium,armitage2017weyl}. Very recently, materials exhibiting band degeneracies along a line in momentum space (termed line node semimetals) have been in the limelight. These line nodes are enabled by a combination of time reversal and inversion symmetries, in principle without requiring additional crystal symmetries. There is an intensive ongoing effort to search and characterize such material candidates. Theoretical efforts, grounded in density functional theory computations, have led to a number of materials being proposed to host such line nodal states~\cite{xie2015new,PhysRevB.92.045108,PhysRevLett.115.036806,PhysRevLett.115.036807,PhysRevLett.115.026403,yamakage2015line,li2016dirac,chan20163,huang2016topological,hirayama2017topological,xu2017topological,quan2017single,geilhufe2017three}. These range from elemental solids~\cite{hirayama2017topological,li2016dirac} to ternary compounds~\cite{PhysRevLett.115.036806,PhysRevLett.115.036807}, to highlight just a few. Promising experimental advances in synthesizing and characterizing candidate materials have also been reported for PtSn$_{4}$~\cite{wu2016dirac}, PbTaSe$_{2}$~\cite{bian2016topological}, ZrSiS~\cite{schoop2016dirac,neupane2016observation,singha2017large}, ZrSiSe~\cite{hu2016evidence}, and CaAgP~\cite{okamoto2016low,emmanouilidou2017magnetotransport}.

Among the systems that have been identified to show this feature in the band structure, alkaline earth metal tripnictides of the form AB$_3$ (A=Ca, Sr and B=P, As) are a promising set~\cite{xu2017topological,quan2017single}. The calcium compounds, CaP$_{3}$ and CaAs$_{3}$ form in the $P\overline{1}$ space group. Intriguingly, the only structural symmetry is an inversion symmetry and the structure lacks any other crystal symmetries. Owing to this minimum symmetry condition, these materials have been dubbed the "hydrogen atom" of line nodal semimetals~\cite{quan2017single}. 

\begin{figure*}
\includegraphics[scale=1.0]{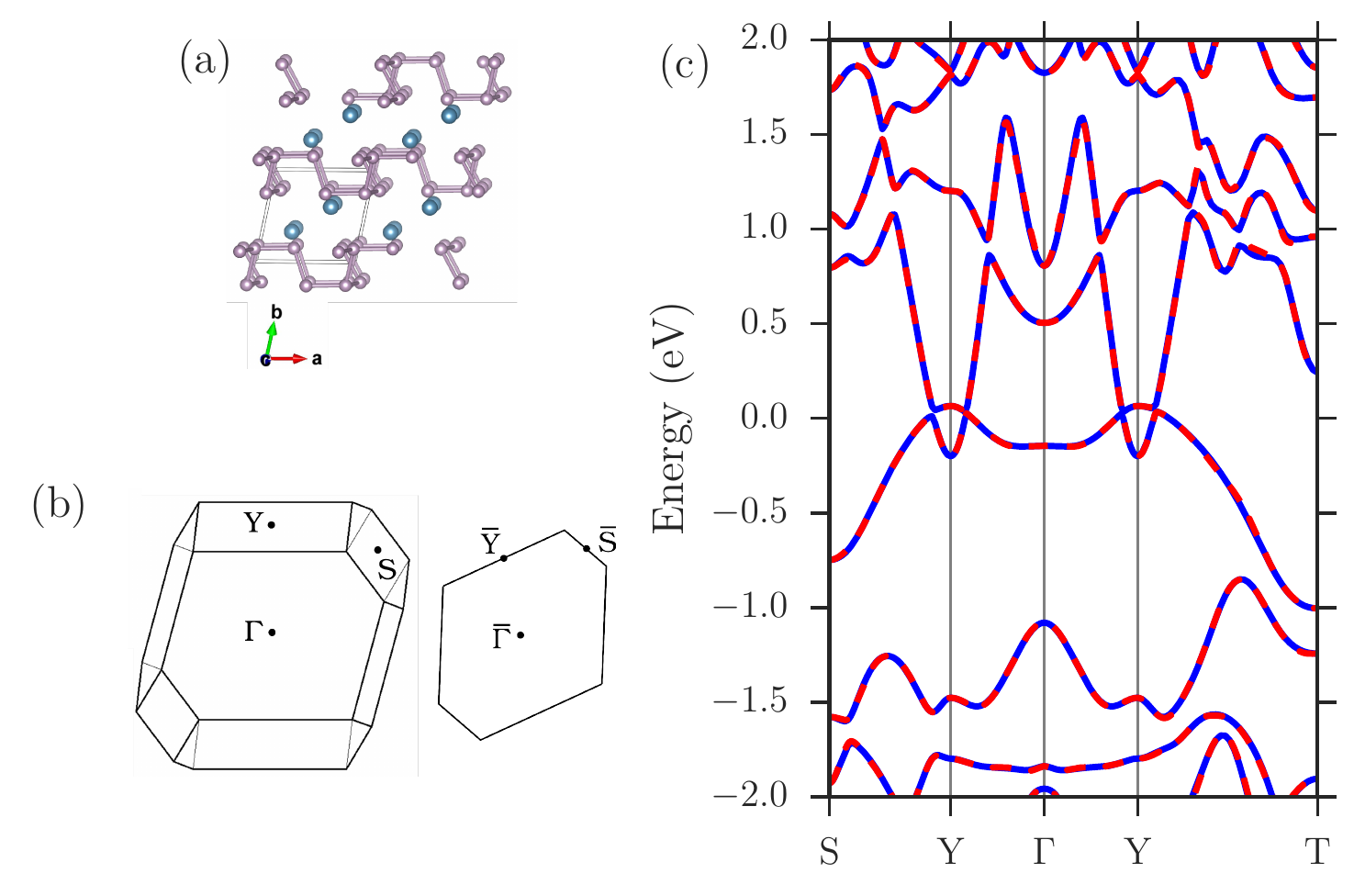}
  \caption{\textbf{Crystal structure and bulk band structure.} (a) Crystal structure of CaP$_3$ viewed along $c$ axis. Larger blue spheres represent Ca atoms, while smaller purple ones indicate P atoms. Solid black lines enclose the eight atom unit cell. (b) Bulk (left) and surface (right) Brillouin zone with the high symmetry points marked. (c) Bulk band structure of CaP$_3$. Solid blue lines are obtained from density functional theory, while the dashed lines are bands obtained after a Wannier projection. Note the gapless features around Y point of the Brillouin zone.}  \label{bulk_bands}
\end{figure*}

In this contribution, we systematically analyze the bulk and finite size effects in prototypical line node material CaP$_{3}$ using density functional theory based computations. Furthermore, we show that tuning the volume of the material, provides a handle on the reciprocal space radius of the line node. It is important to characterize the band structure in confined geometries. The bulk contribution to transport in a suitable slab geometry can be suppressed, facilitating the observation of effects from the surface drumhead states. In the past, this strategy has been proved to be important for topological insulators~\cite{zhang2010crossover}. Moreover, a number of experimental investigations can be performed in the thin film setup, including creation of heterojunctions with other materials. Therefore, it is useful to understand how the band structure of line node materials evolves with thickness. Here, we compute the thickness dependent band structure of prototypical line node material CaP$_{3}$, and estimate the minimum thickness needed to observe the bulk line node. Controlling the position and extent of the bulk line node and the resulting surface states remains an intriguing aspect of these materials. Using ab initio calculations, we show that the line node radius in CaP$_{3}$ is rather sensitive to bulk volume change. Further, we rationalize this observation based on a low energy model. We propose that application of pressure or suitable substrate engineering can be used as an effective means to control and tune the line node and the surface drumhead states. The resulting changes in the band structure could be probed either directly by angle resolved photoemission or via quantum oscillation measurements to map the Fermi surfaces.

\section{Methods}

Density functional theory calculations were performed using the generalized gradient approximation (GGA) to the exchange-correlation functional~\cite{perdew1996generalized}, as implemented in the Vienna Ab-Initio Simulation Package ({\sc vasp})~\cite{kresse1996efficiency,kresse1996efficient}. We used a plane wave cutoff of 500 eV and the bulk Brillouin zone was sampled using a $8\times 8\times 8$ $k$-point mesh. We used experimental coordinates for CaP$_3$ as reported in Ref.~\onlinecite{dahlmann1973cap}. The starting point to obtain quantum oscillation frequencies are the band energies over a dense $51\times 51\times 51$ $k$-point grid. We use the Supercell K-space Extremal Area Finder ({\sc skeaf}) algorithm of Rourke and Julian~\cite{julian2012numerical} to obtain the extremal orbits and the resulting frequencies. In brief, the method proceeds by creating a reciprocal space supercell, which is divided into slices perpendicular to the direction of the applied magnetic field. The obtained orbits for each slice are matched to neighbouring slices, allowing determination of the extremal orbits. In the past, this method has proved useful to analyze and compare with quantum oscillation measurements~\cite{giraldo2016fermi}.

\begin{figure*}
\includegraphics[scale=1.0]{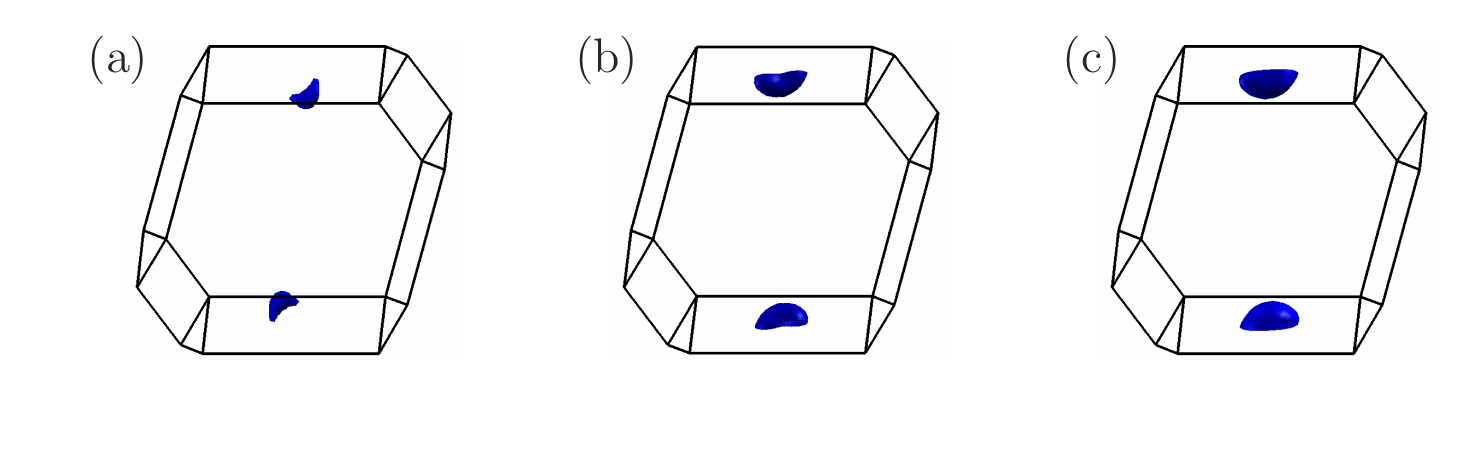}
  \caption{\textbf{Bulk Fermi surface.} Bulk Fermi surface for (a)$E-E_{F}=0$ eV, (b) $E-E_{F}=0.07$ eV, and (c) $E-E_{F}=0.1$ eV. The line node features arise as ``toroidal arcs'' connected across the boundary of the Brillouin zone. With increasing Fermi energy these transform to an ellipsoid.}  \label{fermi_surface}
\end{figure*}

\begin{figure*}
\includegraphics[scale=1.0]{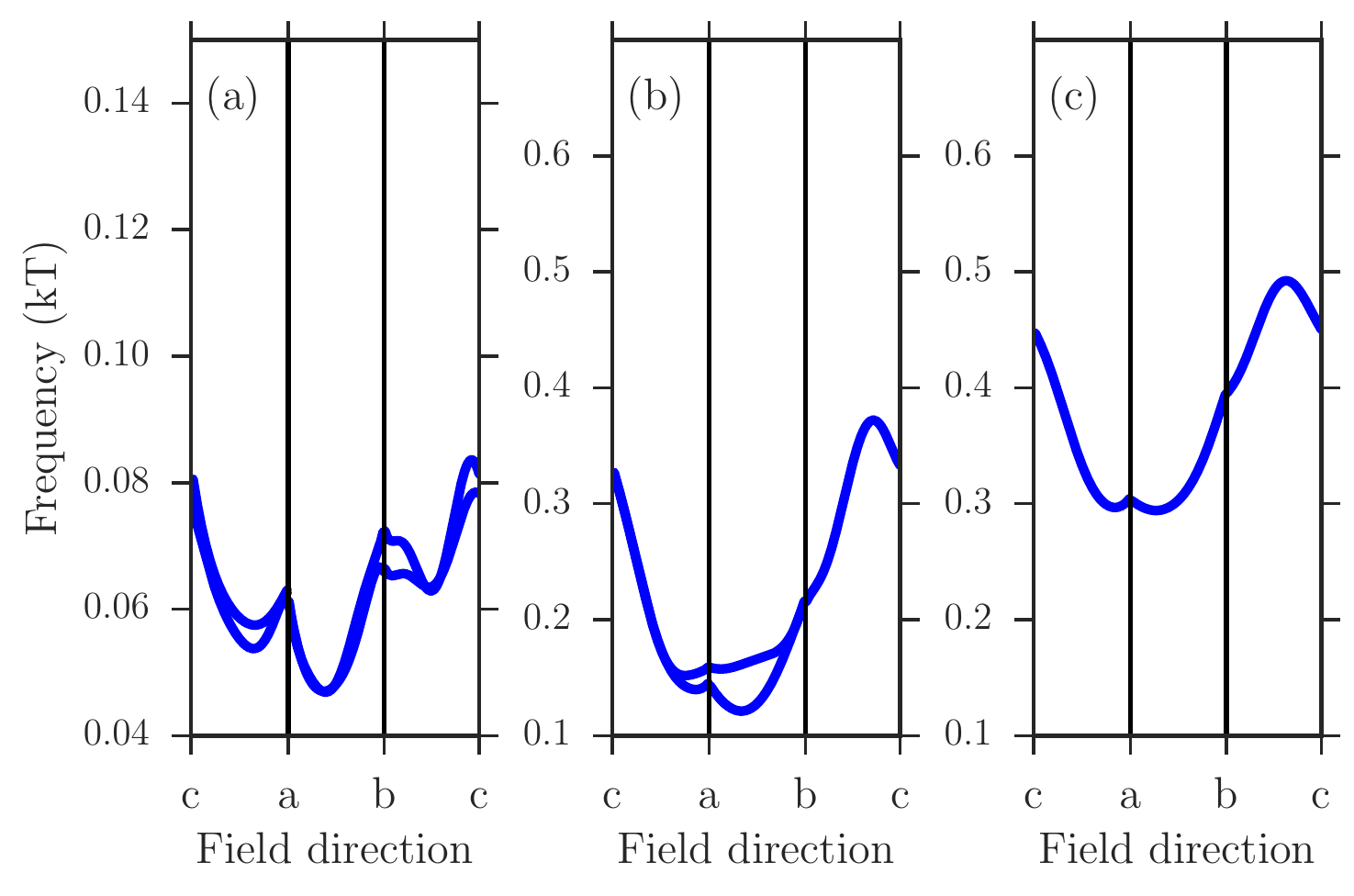}
  \caption{\textbf{Quantum oscillations with increasing Fermi energy.} Quantum oscillation frequencies for different directions of applied magnetic fields with Fermi energy at (a) $E-E_{F}=0$ eV, (b) $E-E_{F}=0.07$ eV, and (c) $E-E_{F}=0.1$ eV. Signature of Lifshitz transition from toroidal to ellipsoidal Fermi surface is seen in degeneracy of the calculated frequencies.}  \label{quantum_oscillations}
\end{figure*}

To investigate slab geometries, we downfolded the band structure obtained from density functional theory to an effective tight-binding model using the $p_{x},p_{y},p_{z}$ orbitals of the pnictide atoms. The downfolding was achieved by constructing maximally localized wannier functions, employing the {\sc wannier90} suite~\cite{mostofi2008wannier90}. Finally, the resulting tight-binding model was analyzed using the WannierTools package~\cite{wu2017wanniertools}. The surface spectral functions, $\rho(E,k)$, in a semi-infinite geometry were computed using an iterative Green's function method. Here $\rho(E,k)=-\mathrm{tr}[G_{00}(E,k)]/\pi$, where $G_{00}(E,k)$ is the Green's function corresponding to the surface layer.

\begin{figure*}
\begin{center}
  \includegraphics[scale=1.0]{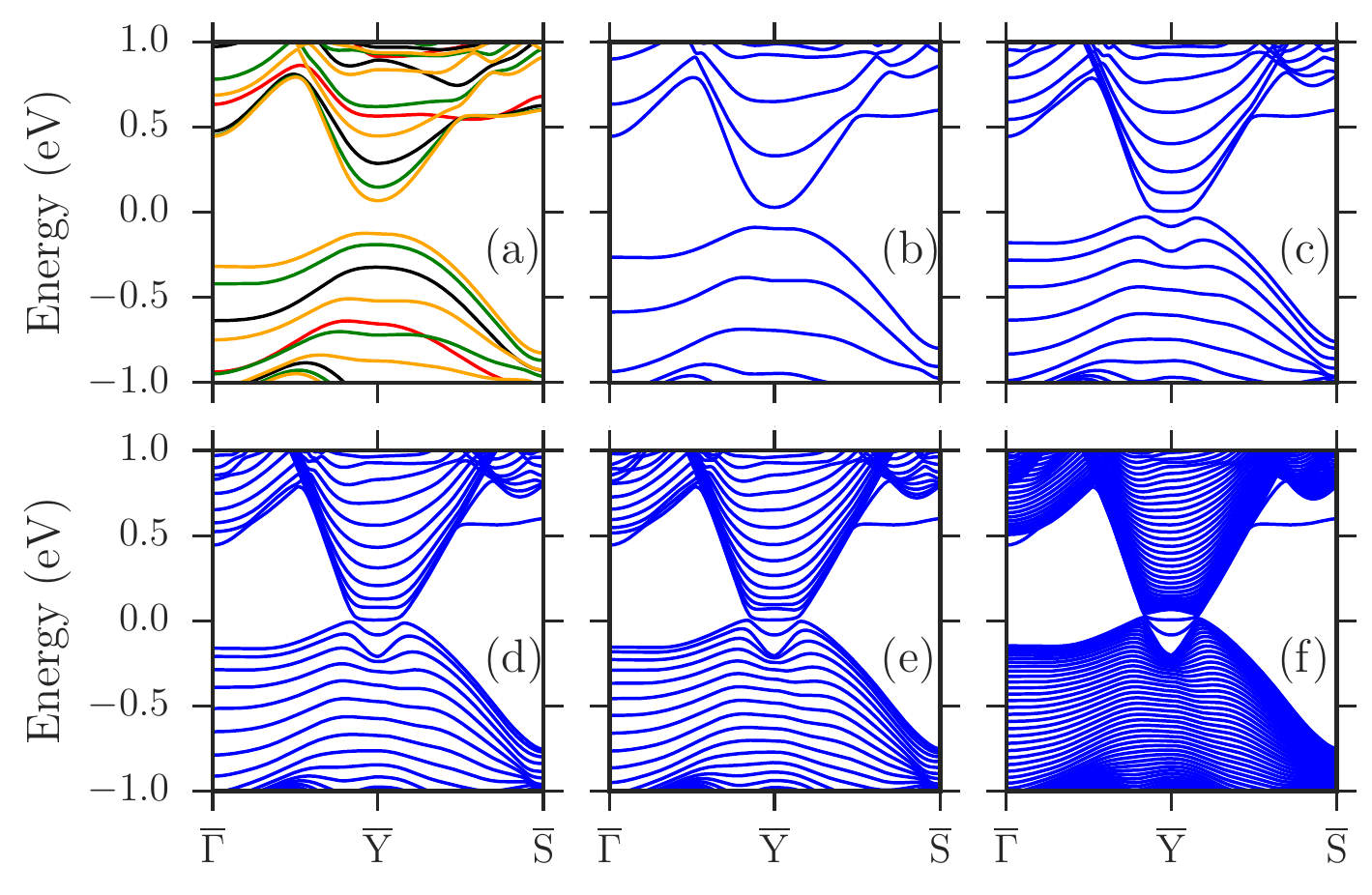}
  \caption{\textbf{Thickness dependence of slab spectrum.} (a) Band structure for ultra thin slabs of thickness 1 layer (red), 2 layers (black), 3 layers (green) and 4 layers (orange). Band structure for slabs of thickness (b) 5 layers, (c) 10 layers, (d) 15 layers, (e) 20 layers, and (f) 50 layers. Note the presence of two "in gap" bands around $\overline{\mathrm{Y}}$ in thicker slabs.}  \label{slab_bands}
\end{center}
\end{figure*}

\begin{figure*}
\begin{center}
  \includegraphics[scale=1.0]{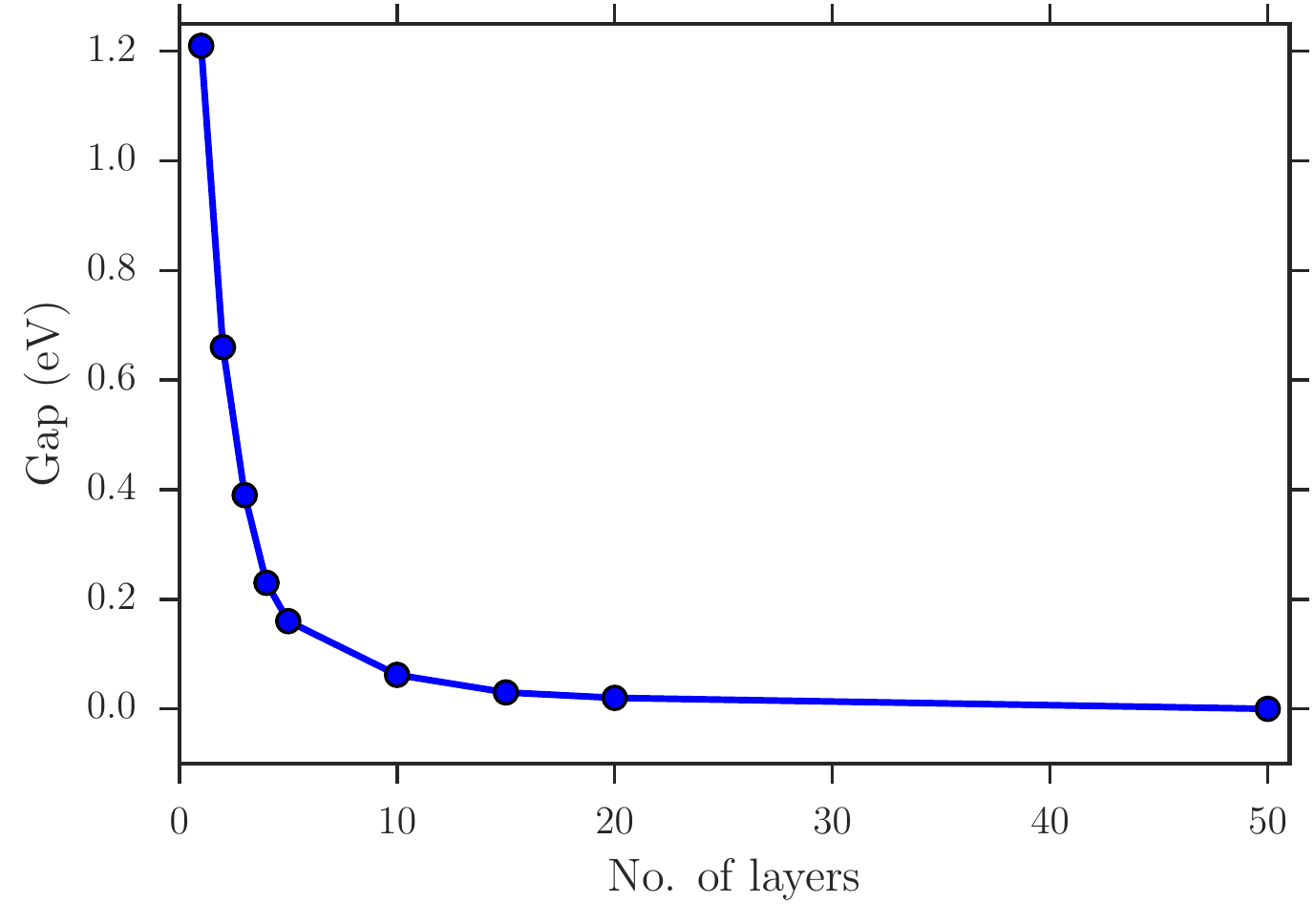}
  \caption{\textbf{Evolution of gap at line nodal point.} Band gap at the $k$-point along $\overline{\Gamma}-\overline{\mathrm{Y}}$, where the line node eventually develops, with increasing slab thickness.}  \label{slab_gap}
\end{center}
\end{figure*}

\begin{figure*}
\begin{center}
  \includegraphics[scale=1.0]{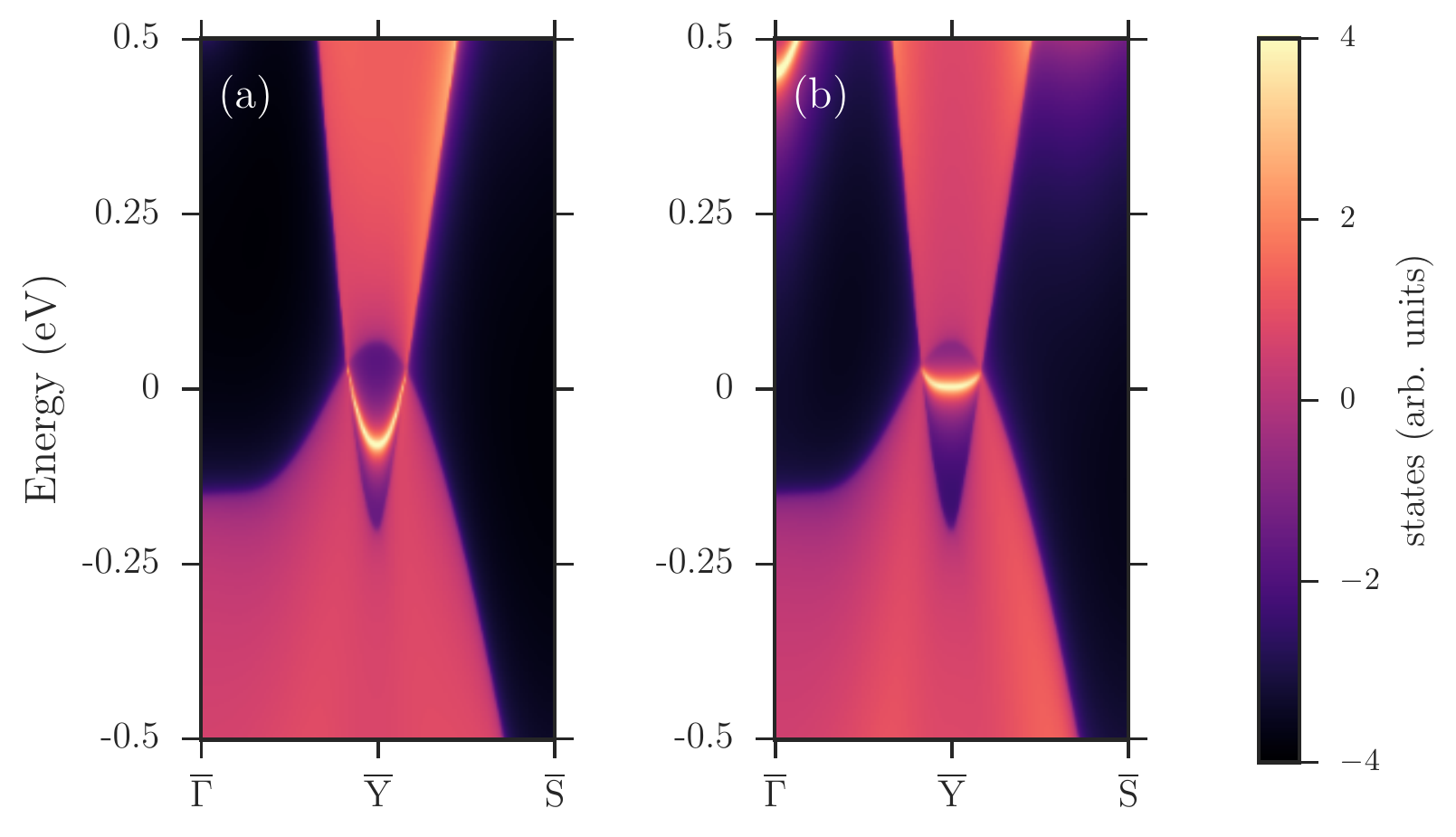}
  \caption{\textbf{Surface spectral function.} Energy dependent surface spectrum for a semi-infinite geometry with (a) Ca termination and (b) P termination. The surface state appears bounded by the bulk line node in both cases.}  \label{surface_dos}
\end{center}
\end{figure*}

\section{Results and Discussion}

Let us begin by discussing the crystal structure of CaP$_{3}$, as shown in Fig.~\ref{bulk_bands}(a). The unit cell (enclosed by black lines) consists of two formula units. The pnictide atoms (denoted by smaller purple spheres) form one dimensional puckered chains, which are separated by the Ca atoms (shown using larger blue spheres). The aforementioned inversion center lies at the center of neighboring Ca atoms. Bulk Brillouin zone and its surface projection are plotted in Fig.~\ref{bulk_bands}(b). The GGA band structure is shown in Fig.~\ref{bulk_bands}(c) using solid lines, with the Fermi energy, $E_{F}$, located at zero. This is in good agreement with previous calculations reported in Ref.~\onlinecite{xu2017topological,quan2017single}. Most notably there is a pair of inverted bands with opposite parity which cross around the $Y$ point of the Brillouin zone. As will be shown subsequently, these form a circular line node. Crucially, these are the only bands around the Fermi level and therefore would allow a direct investigation of the properties of these line nodes without the hindrance of other bands.

Analysis of the atomic contributions of the band structure shows that the bands around the Fermi level predominantly arise from $p_{x},p_{y},p_{z}$ states of P atoms. Therefore, we constructed an effective tight binding model based on these orbitals using a wannierization procedure (as described in the Computational Methods section). The resulting ab initio derived band structure is superimposed on the first-principles result in Fig.~\ref{bulk_bands}(c). Importantly, the two band structures are in good agreement within a wide energy window around the Fermi level signaling the adequacy of the tight binding model.

\begin{figure*}
\begin{center}
  \includegraphics[scale=1.0]{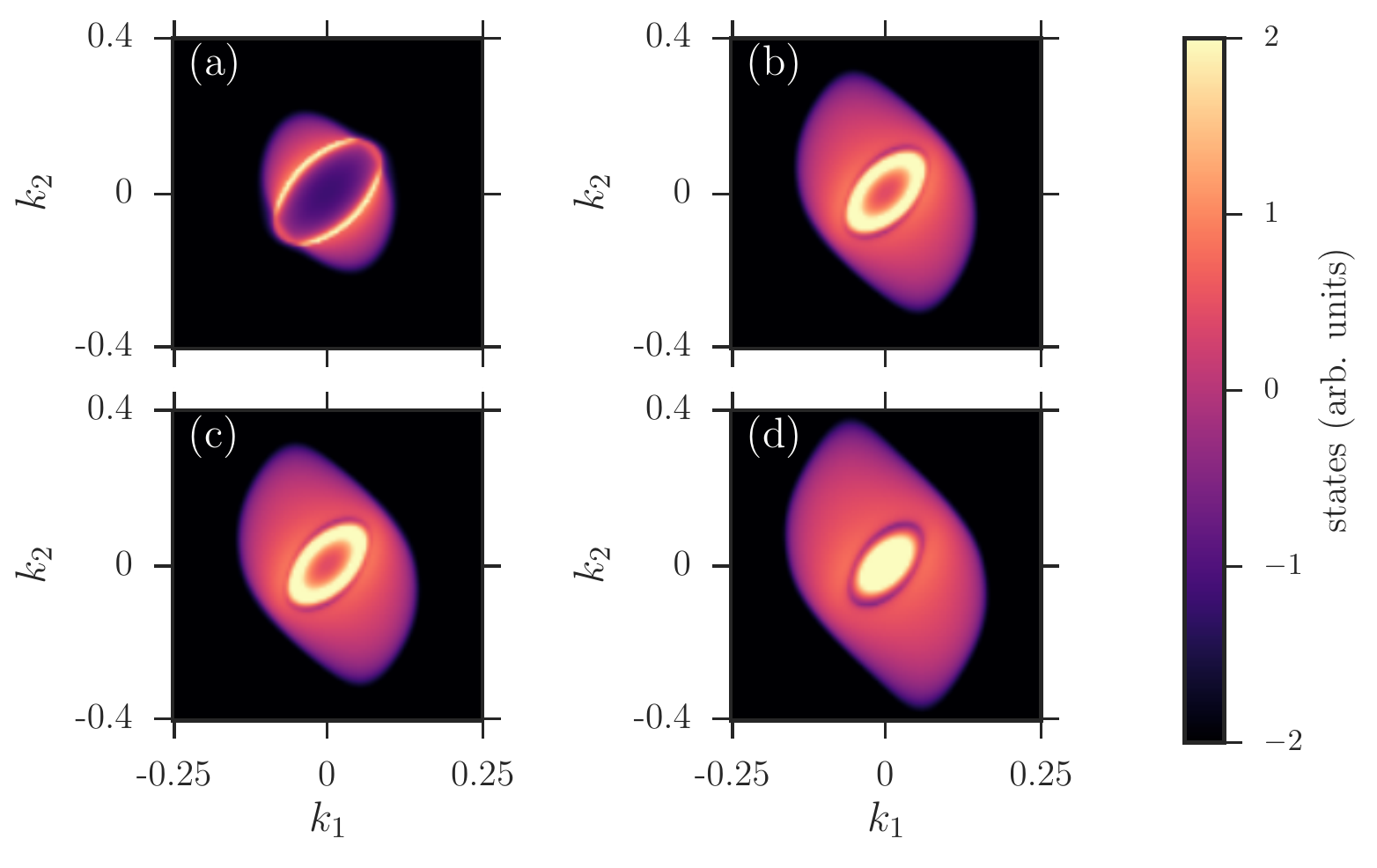}
  \caption{\textbf{Energy dependence of surface state.} Momentum resolved surface spectral function at (a) $E=0$, (b) $E=-0.025$ eV, (c) $E=-0.050$ eV, and (d) $E=-0.075$ eV. $k_1$ and $k_2$ are effective momenta around the center of the line node. The surface state (bright central feature enclosed by the surrounding leaf-like structure) has non-zero dispersion within this 0.1 eV energy window and evolves from arc-like (a) to ring-shaped (b-c) and eventually disk-like (d) as the energy is lowered.}  \label{surface_dos_kx_ky}
\end{center}
\end{figure*}

The line nodal nature of the bands crossing the Fermi level becomes more evident by examining the Fermi surface. These are shown in Fig.~\ref{fermi_surface} for various energies around the Fermi level. Very close to the Fermi level [Fig.~\ref{fermi_surface}(a)], there is a clear loop like structure around the $Y$ point of the Brillouin zone. A closed contour is obtained as the two pieces of the loop connect across the Brillouin zone boundary. The nature and shape of the Fermi surfaces change sharply on both moving slightly above and below the pristine Fermi level. On moving below the pristine Fermi level (not shown), these loop like structures enlarge in size and their shape becomes characteristically toroidal. On the other hand, at a higher Fermi level of 0.1 eV, the Fermi surface undergoes a Lifshitz transition and the Fermi surface acquires cap-like shape around the $Y$ points. These join across the Brillouin zone boundary to form an ellipsoid. We propose that these pronounced changes in the Fermi surface shapes and the substantial electron-hole asymmetry of the band structure can be experimentally probed by means of angle resolved photoemission in doped samples. Another method to see these changes in the Fermi surface topology is through quantum oscillation experiments. We have calculated the oscillation frequencies with changing direction of the applied magnetic field (Fig.~\ref{quantum_oscillations}) using a supercell method to find the extremal orbits (see Computational Methods section for a discussion). Expectedly, the frequencies increase with increasing Fermi level due to the increasing size of the Fermi surface (see Fig.~\ref{fermi_surface} for a visualization). At $E-E_{F}=0$, we find two closely spaced frequencies. These arise from extremal orbits around the thicker and thinner necks of the toroidal Fermi surface. At higher Fermi energy, as the shape of the Fermi surface changes to an ellipsoid only a single frequency is found [Fig.~\ref{quantum_oscillations}(c)]. This corresponds to one extremal orbit traversing the surface of the ellipsoid, whose length varies for different directions of the applied magnetic field. Such a change in the number of oscillation frequencies and their magnitude, with changing Fermi level should be readily observable as a signature of the Lifshitz transition.

Next, let us move on to the thickness dependent electronic structure of CaP$_{3}$. The band structures for the ultra thin slabs (stacked along the crystallographic $c$ direction) of thicknesses ranging from 1 layer to 4 layers are collected in Fig.~\ref{slab_bands}(a). Due to the finite size confinement all these systems are gapped with a band gap of more than 100 meV. The same is true for a 5 layer thick slab [Fig.~\ref{slab_bands}(b)]. The first signatures of the incipient line node starts to be found in 10 layer thick slabs, where a characteristic upturn of the bands around the $\overline{Y}$ point is seen in the bands below the Fermi level, although there is still a gap remaining at these points [Fig.~\ref{slab_bands}(c)]. With subsequent increase in the thickness of the slabs [Fig.~\ref{slab_bands}(d)-(f)], this gap disappears and the line node fully forms. We note here that for thinner slabs having a larger gap, the bands are not inverted around the $\overline{Y}$ point. In contrast, in thicker slabs bands have an inverted order and smaller gaps. In addition, we find two ``in gap'' features within the line nodal circle. These will be examined in more detail subsequently in a semi-infinite setup. 

The evolution of the energy gap, at the $k$-point along the $\overline{\Gamma}-\overline{Y}$ direction where the bulk line node crossing exists, with the thickness of the slab is collected in Fig.~\ref{slab_gap}. There is a rather rapid decay of the gap with thickness initially, which then slows down and reaches to negligible values only above a thickness of about 20 layers. This behavior is rather similar to that seen for prototypical Dirac semimetal materials (i.e., those with a point node degeneracy instead of a line node one)~\cite{narayan2014topological}.

Having examined the electronic structure of CaP$_{3}$ in infinite (i.e. bulk) and finite (i.e. slab) geometries, we next move on to an intermediate analysis in the semi-infinite case. This is particularly relevant for surface sensitive probes employing a thick bulk-like sample, for instance angle resolved photoemission and scanning tunneling spectroscopy. The electronic structure in the semi-infinite geometry is obtained by employing an iterative Green's function technique to calculate the surface spectral functions~\cite{wu2017wanniertools}. The energy dependent surface spectral function is shown in Fig.~\ref{surface_dos} along the high symmetry directions around the $\overline{Y}$ point. The bulk nodal crossings are clearly visible. In addition, there are bright features (corresponding to higher spectral weight on the surface), present within the bulk line node for both terminations of the system. These are the two drumhead states which arise due to the bulk band inversion between the two bands. Notice that this surface band is more dispersive for Ca termination while remains rather flat for the P termination. This difference could be related to different surface potentials that the two terminations provide and suggests the interesting possibility that the drumhead state dispersion could be tuned by surface gating. The effect of the dispersion is seen more clearly by studying the $k$-resolved surface spectral function at different energy values, as shown in Fig.~\ref{surface_dos_kx_ky} for Ca termination. The surface band evolves from an arc-like contour at the Fermi energy [Fig.~\ref{surface_dos_kx_ky}(a)], to a ring-like feature [Fig.~\ref{surface_dos_kx_ky}(b)-(c)] on lowering the energy below the Fermi level. Very close to the bottom of the surface band it appears as a disc around the $\overline{Y}$ point. Such a variation in the surface state dispersion could have implications for comparisons and explanations of future quasi particle interference experiments on these materials. 

\begin{figure*}
\begin{center}
  \includegraphics[scale=1.0]{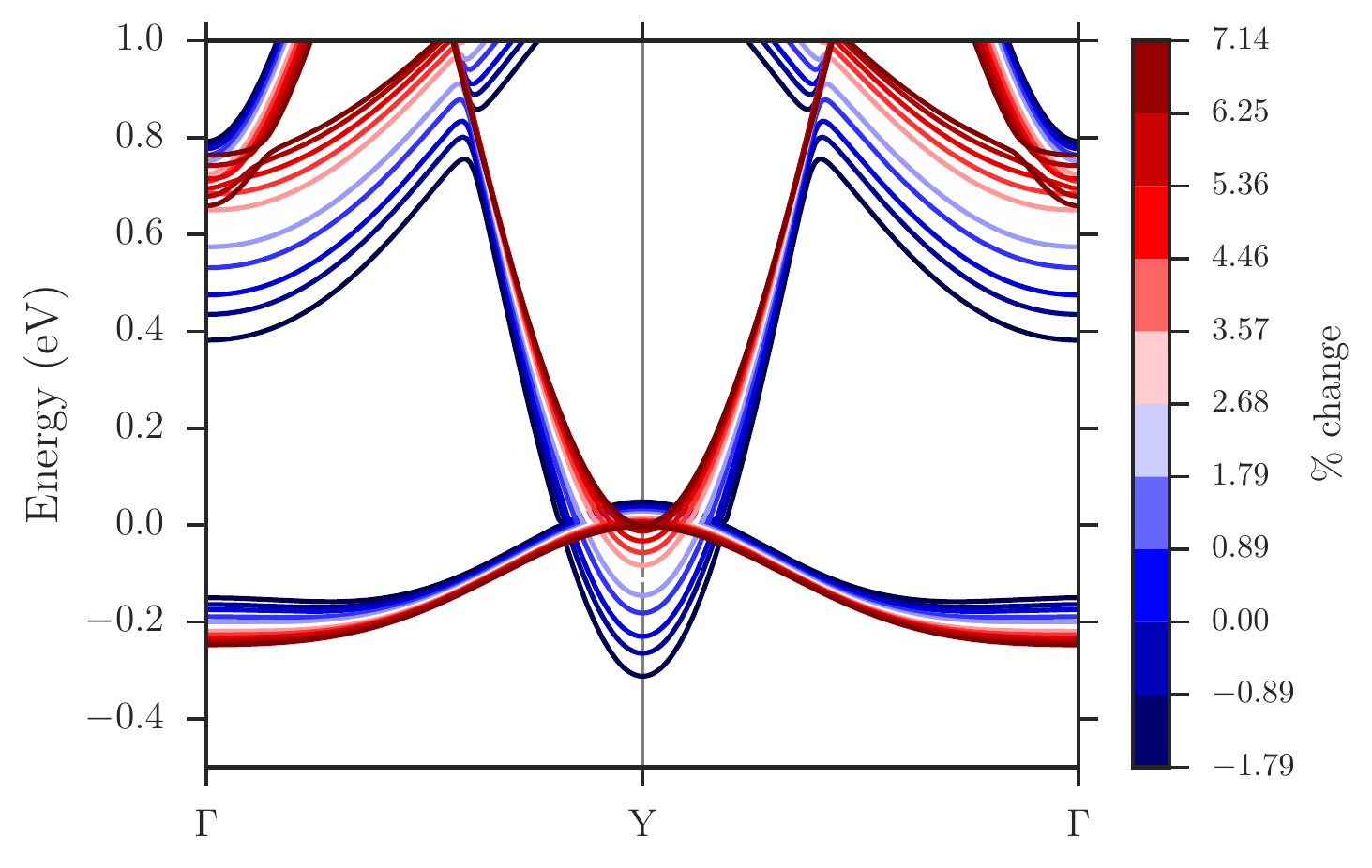}
  \caption{\textbf{Volume tuning of line node.} Evolution of the bulk line node around Y point of the Brillouin zone with varying unit cell volume. With increasing volume the momentum space radius of the line node decreases and so does the "inverted gap" at Y. Such a volume tuning provides a control knob over the radius of the bulk line node and the resulting surface states.}  \label{bulk_bands_pressure}
\end{center}
\end{figure*}

Having investigated the bulk and finite size effects on the electronic structure of the prototypical line node material CaP$_{3}$, we raise the question: how can one control the momentum space extent of the line node and the resulting drumhead states? From our first principles calculations we find that change in volume of the cell has a rather marked effect on the low energy band structure of CaP$_{3}$ (Fig.~\ref{bulk_bands_pressure}). A small change of 1-2\% in the lattice constant already changes the band inversion of the bands forming the line node by close to 50 meV. The effect is specially pronounced on the electron-like band, while the hole-like band is less affected. The result is that the radius of the line node changes with changing volume. An increase in volume reduces the band inversion around the $\overline{Y}$ point, concomitantly reducing the radius of the line node. At rather large values of increased volume the line node eventually transitions to a point node. We note that this large increase in volume needed for the transition in CaP$_{3}$ would be hard to achieve experimentally. However, the results are presented here as a proof of concept and perhaps could be more feasible in other line nodal materials. On the other hand a decrease in the volume increases the band inversion, while at the same time increasing the momentum space radius of the line node. Moderate changes in volume could be practically achieved by either applying pressure or choosing a suitable substrate to tune the volume of the material. The resulting changes in the band structure could then be measured either directly by means of angle resolved photoemission techniques or by monitoring the change in quantum oscillation frequencies. 

To rationalize this observation, let us consider a low-energy Hamiltonian to describe the line node semimetals~\cite{PhysRevLett.115.036806} 

\begin{eqnarray}
 H&=&[\epsilon_{0}+a_{xy}(k_{x}^{2}+k_{y}^{2})+a_{z}k_{z}^{2}]I + vk_{z}\sigma_{y} \nonumber \\
  &+& [\Delta\epsilon+b_{xy}(k_{x}^{2}+k_{y}^{2})+b_{z}k_{z}^{2}]\sigma_{z},
\end{eqnarray}

\noindent where $\sigma_{x},\sigma_{y},\sigma_{z}$ are Pauli matrices and $I$ is the $2\times 2$ identity matrix. For this model, we have symmetry operations inversion, $\mathcal{P}=\sigma_{z}$ and time reversal, $\mathcal{T}=\mathcal{K}$, where $\mathcal{K}$ denotes complex conjugation. The energy eigenvalues for this model are given by $E=\epsilon_{0}+a_{xy}(k_{x}^{2}+k_{y}^{2})+a_{z}k_{z}^{2} \pm \sqrt{[\Delta\epsilon+b_{xy}(k_{x}^{2}+k_{y}^{2})+b_{z}k_{z}^{2}]^{2}+v^{2}k_{z}^{2}}$. For $\Delta\epsilon <0$, i.e. a finite band inversion, the two bands form a line node degeneracy at $k_{z}=0$ and $k_{x}^{2} + k_{y}^{2} = -\Delta\epsilon/b_{xy}$. We notice that the radius of the line node is proportional to the band inversion strength $\Delta\epsilon$, and that increasing band inversion strength increases the radius of the line node and vice versa. When the band inversion strength vanishes ($\Delta\epsilon =0$), the system is gapless with a point node, i.e., the line node radius is zero. This analysis matches our observations from density functional theory calculations. Further, this suggests that while our calculations are for specific material CaP$_{3}$, the present strategy for controlling the line node could prove to be useful for other line nodal systems.

\section{Conclusions}

To summarize, we systematically analyzed the bulk and finite size effects on the band structure of prototypical line node material CaP$_{3}$ using density functional theory based computations. For the bulk system, we analyzed quantum oscillations with changes in the Fermi surface topology. We showed that ultra thin slabs are gapped, with the first signatures of the line node appearing around a thickness of ten layers. We showed that the line node radius in CaP$_{3}$ is rather sensitive to bulk volume change, which we also rationalized using an effective low energy model. Based on this observation, we proposed that application of pressure or suitable substrate engineering can be used as an effective means to control and tune the line node and the accompanying surface drumhead states. These changes could then be experimentally probed using either angle resolved photoemission techniques or by employing quantum oscillations to map out the Fermi surfaces.

\section*{Acknowledgments}

I would like to thank B. Sangiorgio for insightful discussions, C. Setty for a related collaboration, and N. Spaldin for valuable advice. Support from ETH Zurich is gratefully acknowledged.

\bibliography{references}

\begin{thebibliography}{35}
\expandafter\ifx\csname natexlab\endcsname\relax\def\natexlab#1{#1}\fi
\expandafter\ifx\csname bibnamefont\endcsname\relax
  \def\bibnamefont#1{#1}\fi
\expandafter\ifx\csname bibfnamefont\endcsname\relax
  \def\bibfnamefont#1{#1}\fi
\expandafter\ifx\csname citenamefont\endcsname\relax
  \def\citenamefont#1{#1}\fi
\expandafter\ifx\csname url\endcsname\relax
  \def\url#1{\texttt{#1}}\fi
\expandafter\ifx\csname urlprefix\endcsname\relax\def\urlprefix{URL }\fi
\providecommand{\bibinfo}[2]{#2}
\providecommand{\eprint}[2][]{\url{#2}}

\bibitem[{\citenamefont{Hasan and Kane}(2010)}]{hasan2010colloquium}
\bibinfo{author}{\bibfnamefont{M.~Z.} \bibnamefont{Hasan}} \bibnamefont{and}
  \bibinfo{author}{\bibfnamefont{C.~L.} \bibnamefont{Kane}},
  \bibinfo{journal}{Reviews of Modern Physics} \textbf{\bibinfo{volume}{82}},
  \bibinfo{pages}{3045} (\bibinfo{year}{2010}).

\bibitem[{\citenamefont{Qi and Zhang}(2011)}]{qi2011topological}
\bibinfo{author}{\bibfnamefont{X.-L.} \bibnamefont{Qi}} \bibnamefont{and}
  \bibinfo{author}{\bibfnamefont{S.-C.} \bibnamefont{Zhang}},
  \bibinfo{journal}{Reviews of Modern Physics} \textbf{\bibinfo{volume}{83}},
  \bibinfo{pages}{1057} (\bibinfo{year}{2011}).

\bibitem[{\citenamefont{Bansil et~al.}(2016)\citenamefont{Bansil, Lin, and
  Das}}]{bansil2016colloquium}
\bibinfo{author}{\bibfnamefont{A.}~\bibnamefont{Bansil}},
  \bibinfo{author}{\bibfnamefont{H.}~\bibnamefont{Lin}}, \bibnamefont{and}
  \bibinfo{author}{\bibfnamefont{T.}~\bibnamefont{Das}},
  \bibinfo{journal}{Reviews of Modern Physics} \textbf{\bibinfo{volume}{88}},
  \bibinfo{pages}{021004} (\bibinfo{year}{2016}).

\bibitem[{\citenamefont{Armitage et~al.}(2017)\citenamefont{Armitage, Mele, and
  Vishwanath}}]{armitage2017weyl}
\bibinfo{author}{\bibfnamefont{N.}~\bibnamefont{Armitage}},
  \bibinfo{author}{\bibfnamefont{E.}~\bibnamefont{Mele}}, \bibnamefont{and}
  \bibinfo{author}{\bibfnamefont{A.}~\bibnamefont{Vishwanath}},
  \bibinfo{journal}{arXiv preprint arXiv:1705.01111}  (\bibinfo{year}{2017}).

\bibitem[{\citenamefont{Xie et~al.}(2015)\citenamefont{Xie, Schoop, Seibel,
  Gibson, Xie, and Cava}}]{xie2015new}
\bibinfo{author}{\bibfnamefont{L.~S.} \bibnamefont{Xie}},
  \bibinfo{author}{\bibfnamefont{L.~M.} \bibnamefont{Schoop}},
  \bibinfo{author}{\bibfnamefont{E.~M.} \bibnamefont{Seibel}},
  \bibinfo{author}{\bibfnamefont{Q.~D.} \bibnamefont{Gibson}},
  \bibinfo{author}{\bibfnamefont{W.}~\bibnamefont{Xie}}, \bibnamefont{and}
  \bibinfo{author}{\bibfnamefont{R.~J.} \bibnamefont{Cava}},
  \bibinfo{journal}{APL Materials} \textbf{\bibinfo{volume}{3}},
  \bibinfo{pages}{083602} (\bibinfo{year}{2015}).

\bibitem[{\citenamefont{Weng et~al.}(2015)\citenamefont{Weng, Liang, Xu, Yu,
  Fang, Dai, and Kawazoe}}]{PhysRevB.92.045108}
\bibinfo{author}{\bibfnamefont{H.}~\bibnamefont{Weng}},
  \bibinfo{author}{\bibfnamefont{Y.}~\bibnamefont{Liang}},
  \bibinfo{author}{\bibfnamefont{Q.}~\bibnamefont{Xu}},
  \bibinfo{author}{\bibfnamefont{R.}~\bibnamefont{Yu}},
  \bibinfo{author}{\bibfnamefont{Z.}~\bibnamefont{Fang}},
  \bibinfo{author}{\bibfnamefont{X.}~\bibnamefont{Dai}}, \bibnamefont{and}
  \bibinfo{author}{\bibfnamefont{Y.}~\bibnamefont{Kawazoe}},
  \bibinfo{journal}{Phys. Rev. B} \textbf{\bibinfo{volume}{92}},
  \bibinfo{pages}{045108} (\bibinfo{year}{2015}).

\bibitem[{\citenamefont{Kim et~al.}(2015)\citenamefont{Kim, Wieder, Kane, and
  Rappe}}]{PhysRevLett.115.036806}
\bibinfo{author}{\bibfnamefont{Y.}~\bibnamefont{Kim}},
  \bibinfo{author}{\bibfnamefont{B.~J.} \bibnamefont{Wieder}},
  \bibinfo{author}{\bibfnamefont{C.~L.} \bibnamefont{Kane}}, \bibnamefont{and}
  \bibinfo{author}{\bibfnamefont{A.~M.} \bibnamefont{Rappe}},
  \bibinfo{journal}{Phys. Rev. Lett.} \textbf{\bibinfo{volume}{115}},
  \bibinfo{pages}{036806} (\bibinfo{year}{2015}).

\bibitem[{\citenamefont{Yu et~al.}(2015)\citenamefont{Yu, Weng, Fang, Dai, and
  Hu}}]{PhysRevLett.115.036807}
\bibinfo{author}{\bibfnamefont{R.}~\bibnamefont{Yu}},
  \bibinfo{author}{\bibfnamefont{H.}~\bibnamefont{Weng}},
  \bibinfo{author}{\bibfnamefont{Z.}~\bibnamefont{Fang}},
  \bibinfo{author}{\bibfnamefont{X.}~\bibnamefont{Dai}}, \bibnamefont{and}
  \bibinfo{author}{\bibfnamefont{X.}~\bibnamefont{Hu}}, \bibinfo{journal}{Phys.
  Rev. Lett.} \textbf{\bibinfo{volume}{115}}, \bibinfo{pages}{036807}
  (\bibinfo{year}{2015}).

\bibitem[{\citenamefont{Mullen et~al.}(2015)\citenamefont{Mullen, Uchoa, and
  Glatzhofer}}]{PhysRevLett.115.026403}
\bibinfo{author}{\bibfnamefont{K.}~\bibnamefont{Mullen}},
  \bibinfo{author}{\bibfnamefont{B.}~\bibnamefont{Uchoa}}, \bibnamefont{and}
  \bibinfo{author}{\bibfnamefont{D.~T.} \bibnamefont{Glatzhofer}},
  \bibinfo{journal}{Phys. Rev. Lett.} \textbf{\bibinfo{volume}{115}},
  \bibinfo{pages}{026403} (\bibinfo{year}{2015}).

\bibitem[{\citenamefont{Yamakage et~al.}(2015)\citenamefont{Yamakage, Yamakawa,
  Tanaka, and Okamoto}}]{yamakage2015line}
\bibinfo{author}{\bibfnamefont{A.}~\bibnamefont{Yamakage}},
  \bibinfo{author}{\bibfnamefont{Y.}~\bibnamefont{Yamakawa}},
  \bibinfo{author}{\bibfnamefont{Y.}~\bibnamefont{Tanaka}}, \bibnamefont{and}
  \bibinfo{author}{\bibfnamefont{Y.}~\bibnamefont{Okamoto}},
  \bibinfo{journal}{Journal of the Physical Society of Japan}
  \textbf{\bibinfo{volume}{85}}, \bibinfo{pages}{013708}
  (\bibinfo{year}{2015}).

\bibitem[{\citenamefont{Li et~al.}(2016)\citenamefont{Li, Ma, Cheng, Wang, Li,
  Zhang, Li, and Chen}}]{li2016dirac}
\bibinfo{author}{\bibfnamefont{R.}~\bibnamefont{Li}},
  \bibinfo{author}{\bibfnamefont{H.}~\bibnamefont{Ma}},
  \bibinfo{author}{\bibfnamefont{X.}~\bibnamefont{Cheng}},
  \bibinfo{author}{\bibfnamefont{S.}~\bibnamefont{Wang}},
  \bibinfo{author}{\bibfnamefont{D.}~\bibnamefont{Li}},
  \bibinfo{author}{\bibfnamefont{Z.}~\bibnamefont{Zhang}},
  \bibinfo{author}{\bibfnamefont{Y.}~\bibnamefont{Li}}, \bibnamefont{and}
  \bibinfo{author}{\bibfnamefont{X.-Q.} \bibnamefont{Chen}},
  \bibinfo{journal}{Physical review letters} \textbf{\bibinfo{volume}{117}},
  \bibinfo{pages}{096401} (\bibinfo{year}{2016}).

\bibitem[{\citenamefont{Chan et~al.}(2016)\citenamefont{Chan, Chiu, Chou, and
  Schnyder}}]{chan20163}
\bibinfo{author}{\bibfnamefont{Y.-H.} \bibnamefont{Chan}},
  \bibinfo{author}{\bibfnamefont{C.-K.} \bibnamefont{Chiu}},
  \bibinfo{author}{\bibfnamefont{M.}~\bibnamefont{Chou}}, \bibnamefont{and}
  \bibinfo{author}{\bibfnamefont{A.~P.} \bibnamefont{Schnyder}},
  \bibinfo{journal}{Physical Review B} \textbf{\bibinfo{volume}{93}},
  \bibinfo{pages}{205132} (\bibinfo{year}{2016}).

\bibitem[{\citenamefont{Huang et~al.}(2016)\citenamefont{Huang, Liu,
  Vanderbilt, and Duan}}]{huang2016topological}
\bibinfo{author}{\bibfnamefont{H.}~\bibnamefont{Huang}},
  \bibinfo{author}{\bibfnamefont{J.}~\bibnamefont{Liu}},
  \bibinfo{author}{\bibfnamefont{D.}~\bibnamefont{Vanderbilt}},
  \bibnamefont{and} \bibinfo{author}{\bibfnamefont{W.}~\bibnamefont{Duan}},
  \bibinfo{journal}{Physical Review B} \textbf{\bibinfo{volume}{93}},
  \bibinfo{pages}{201114} (\bibinfo{year}{2016}).

\bibitem[{\citenamefont{Hirayama et~al.}(2017)\citenamefont{Hirayama, Okugawa,
  Miyake, and Murakami}}]{hirayama2017topological}
\bibinfo{author}{\bibfnamefont{M.}~\bibnamefont{Hirayama}},
  \bibinfo{author}{\bibfnamefont{R.}~\bibnamefont{Okugawa}},
  \bibinfo{author}{\bibfnamefont{T.}~\bibnamefont{Miyake}}, \bibnamefont{and}
  \bibinfo{author}{\bibfnamefont{S.}~\bibnamefont{Murakami}},
  \bibinfo{journal}{Nature communications} \textbf{\bibinfo{volume}{8}}
  (\bibinfo{year}{2017}).

\bibitem[{\citenamefont{Xu et~al.}(2017)\citenamefont{Xu, Yu, Fang, Dai, and
  Weng}}]{xu2017topological}
\bibinfo{author}{\bibfnamefont{Q.}~\bibnamefont{Xu}},
  \bibinfo{author}{\bibfnamefont{R.}~\bibnamefont{Yu}},
  \bibinfo{author}{\bibfnamefont{Z.}~\bibnamefont{Fang}},
  \bibinfo{author}{\bibfnamefont{X.}~\bibnamefont{Dai}}, \bibnamefont{and}
  \bibinfo{author}{\bibfnamefont{H.}~\bibnamefont{Weng}},
  \bibinfo{journal}{Physical Review B} \textbf{\bibinfo{volume}{95}},
  \bibinfo{pages}{045136} (\bibinfo{year}{2017}).

\bibitem[{\citenamefont{Quan et~al.}(2017)\citenamefont{Quan, Yin, and
  Pickett}}]{quan2017single}
\bibinfo{author}{\bibfnamefont{Y.}~\bibnamefont{Quan}},
  \bibinfo{author}{\bibfnamefont{Z.}~\bibnamefont{Yin}}, \bibnamefont{and}
  \bibinfo{author}{\bibfnamefont{W.}~\bibnamefont{Pickett}},
  \bibinfo{journal}{Physical Review Letters} \textbf{\bibinfo{volume}{118}},
  \bibinfo{pages}{176402} (\bibinfo{year}{2017}).

\bibitem[{\citenamefont{Geilhufe et~al.}(2017)\citenamefont{Geilhufe, Bouhon,
  Borysov, and Balatsky}}]{geilhufe2017three}
\bibinfo{author}{\bibfnamefont{R.~M.} \bibnamefont{Geilhufe}},
  \bibinfo{author}{\bibfnamefont{A.}~\bibnamefont{Bouhon}},
  \bibinfo{author}{\bibfnamefont{S.~S.} \bibnamefont{Borysov}},
  \bibnamefont{and} \bibinfo{author}{\bibfnamefont{A.~V.}
  \bibnamefont{Balatsky}}, \bibinfo{journal}{Physical Review B}
  \textbf{\bibinfo{volume}{95}}, \bibinfo{pages}{041103}
  (\bibinfo{year}{2017}).

\bibitem[{\citenamefont{Wu et~al.}(2016)\citenamefont{Wu, Wang, Mun, Johnson,
  Mou, Huang, Lee, Bud’ko, Canfield, and Kaminski}}]{wu2016dirac}
\bibinfo{author}{\bibfnamefont{Y.}~\bibnamefont{Wu}},
  \bibinfo{author}{\bibfnamefont{L.-L.} \bibnamefont{Wang}},
  \bibinfo{author}{\bibfnamefont{E.}~\bibnamefont{Mun}},
  \bibinfo{author}{\bibfnamefont{D.}~\bibnamefont{Johnson}},
  \bibinfo{author}{\bibfnamefont{D.}~\bibnamefont{Mou}},
  \bibinfo{author}{\bibfnamefont{L.}~\bibnamefont{Huang}},
  \bibinfo{author}{\bibfnamefont{Y.}~\bibnamefont{Lee}},
  \bibinfo{author}{\bibfnamefont{S.}~\bibnamefont{Bud’ko}},
  \bibinfo{author}{\bibfnamefont{P.}~\bibnamefont{Canfield}}, \bibnamefont{and}
  \bibinfo{author}{\bibfnamefont{A.}~\bibnamefont{Kaminski}},
  \bibinfo{journal}{Nature Physics}  (\bibinfo{year}{2016}).

\bibitem[{\citenamefont{Bian et~al.}(2016)\citenamefont{Bian, Chang, Sankar,
  Xu, Zheng, Neupert, Chiu, Huang, Chang, Belopolski
  et~al.}}]{bian2016topological}
\bibinfo{author}{\bibfnamefont{G.}~\bibnamefont{Bian}},
  \bibinfo{author}{\bibfnamefont{T.-R.} \bibnamefont{Chang}},
  \bibinfo{author}{\bibfnamefont{R.}~\bibnamefont{Sankar}},
  \bibinfo{author}{\bibfnamefont{S.-Y.} \bibnamefont{Xu}},
  \bibinfo{author}{\bibfnamefont{H.}~\bibnamefont{Zheng}},
  \bibinfo{author}{\bibfnamefont{T.}~\bibnamefont{Neupert}},
  \bibinfo{author}{\bibfnamefont{C.-K.} \bibnamefont{Chiu}},
  \bibinfo{author}{\bibfnamefont{S.-M.} \bibnamefont{Huang}},
  \bibinfo{author}{\bibfnamefont{G.}~\bibnamefont{Chang}},
  \bibinfo{author}{\bibfnamefont{I.}~\bibnamefont{Belopolski}},
  \bibnamefont{et~al.}, \bibinfo{journal}{Nature communications}
  \textbf{\bibinfo{volume}{7}} (\bibinfo{year}{2016}).

\bibitem[{\citenamefont{Schoop et~al.}(2016)\citenamefont{Schoop, Ali,
  Stra{\ss}er, Topp, Varykhalov, Marchenko, Duppel, Parkin, Lotsch, and
  Ast}}]{schoop2016dirac}
\bibinfo{author}{\bibfnamefont{L.~M.} \bibnamefont{Schoop}},
  \bibinfo{author}{\bibfnamefont{M.~N.} \bibnamefont{Ali}},
  \bibinfo{author}{\bibfnamefont{C.}~\bibnamefont{Stra{\ss}er}},
  \bibinfo{author}{\bibfnamefont{A.}~\bibnamefont{Topp}},
  \bibinfo{author}{\bibfnamefont{A.}~\bibnamefont{Varykhalov}},
  \bibinfo{author}{\bibfnamefont{D.}~\bibnamefont{Marchenko}},
  \bibinfo{author}{\bibfnamefont{V.}~\bibnamefont{Duppel}},
  \bibinfo{author}{\bibfnamefont{S.~S.} \bibnamefont{Parkin}},
  \bibinfo{author}{\bibfnamefont{B.~V.} \bibnamefont{Lotsch}},
  \bibnamefont{and} \bibinfo{author}{\bibfnamefont{C.~R.} \bibnamefont{Ast}},
  \bibinfo{journal}{Nature communications} \textbf{\bibinfo{volume}{7}}
  (\bibinfo{year}{2016}).

\bibitem[{\citenamefont{Neupane et~al.}(2016)\citenamefont{Neupane, Belopolski,
  Hosen, Sanchez, Sankar, Szlawska, Xu, Dimitri, Dhakal, Maldonado
  et~al.}}]{neupane2016observation}
\bibinfo{author}{\bibfnamefont{M.}~\bibnamefont{Neupane}},
  \bibinfo{author}{\bibfnamefont{I.}~\bibnamefont{Belopolski}},
  \bibinfo{author}{\bibfnamefont{M.~M.} \bibnamefont{Hosen}},
  \bibinfo{author}{\bibfnamefont{D.~S.} \bibnamefont{Sanchez}},
  \bibinfo{author}{\bibfnamefont{R.}~\bibnamefont{Sankar}},
  \bibinfo{author}{\bibfnamefont{M.}~\bibnamefont{Szlawska}},
  \bibinfo{author}{\bibfnamefont{S.-Y.} \bibnamefont{Xu}},
  \bibinfo{author}{\bibfnamefont{K.}~\bibnamefont{Dimitri}},
  \bibinfo{author}{\bibfnamefont{N.}~\bibnamefont{Dhakal}},
  \bibinfo{author}{\bibfnamefont{P.}~\bibnamefont{Maldonado}},
  \bibnamefont{et~al.}, \bibinfo{journal}{Physical Review B}
  \textbf{\bibinfo{volume}{93}}, \bibinfo{pages}{201104}
  (\bibinfo{year}{2016}).

\bibitem[{\citenamefont{Singha et~al.}(2017)\citenamefont{Singha, Pariari,
  Satpati, and Mandal}}]{singha2017large}
\bibinfo{author}{\bibfnamefont{R.}~\bibnamefont{Singha}},
  \bibinfo{author}{\bibfnamefont{A.~K.} \bibnamefont{Pariari}},
  \bibinfo{author}{\bibfnamefont{B.}~\bibnamefont{Satpati}}, \bibnamefont{and}
  \bibinfo{author}{\bibfnamefont{P.}~\bibnamefont{Mandal}},
  \bibinfo{journal}{Proceedings of the National Academy of Sciences}
  \textbf{\bibinfo{volume}{114}}, \bibinfo{pages}{2468} (\bibinfo{year}{2017}).

\bibitem[{\citenamefont{Hu et~al.}(2016)\citenamefont{Hu, Tang, Liu, Liu, Zhu,
  Graf, Myhro, Tran, Lau, Wei et~al.}}]{hu2016evidence}
\bibinfo{author}{\bibfnamefont{J.}~\bibnamefont{Hu}},
  \bibinfo{author}{\bibfnamefont{Z.}~\bibnamefont{Tang}},
  \bibinfo{author}{\bibfnamefont{J.}~\bibnamefont{Liu}},
  \bibinfo{author}{\bibfnamefont{X.}~\bibnamefont{Liu}},
  \bibinfo{author}{\bibfnamefont{Y.}~\bibnamefont{Zhu}},
  \bibinfo{author}{\bibfnamefont{D.}~\bibnamefont{Graf}},
  \bibinfo{author}{\bibfnamefont{K.}~\bibnamefont{Myhro}},
  \bibinfo{author}{\bibfnamefont{S.}~\bibnamefont{Tran}},
  \bibinfo{author}{\bibfnamefont{C.~N.} \bibnamefont{Lau}},
  \bibinfo{author}{\bibfnamefont{J.}~\bibnamefont{Wei}}, \bibnamefont{et~al.},
  \bibinfo{journal}{Physical review letters} \textbf{\bibinfo{volume}{117}},
  \bibinfo{pages}{016602} (\bibinfo{year}{2016}).

\bibitem[{\citenamefont{Okamoto et~al.}(2016)\citenamefont{Okamoto, Inohara,
  Yamakage, Yamakawa, and Takenaka}}]{okamoto2016low}
\bibinfo{author}{\bibfnamefont{Y.}~\bibnamefont{Okamoto}},
  \bibinfo{author}{\bibfnamefont{T.}~\bibnamefont{Inohara}},
  \bibinfo{author}{\bibfnamefont{A.}~\bibnamefont{Yamakage}},
  \bibinfo{author}{\bibfnamefont{Y.}~\bibnamefont{Yamakawa}}, \bibnamefont{and}
  \bibinfo{author}{\bibfnamefont{K.}~\bibnamefont{Takenaka}},
  \bibinfo{journal}{Journal of the Physical Society of Japan}
  \textbf{\bibinfo{volume}{85}}, \bibinfo{pages}{123701}
  (\bibinfo{year}{2016}).

\bibitem[{\citenamefont{Emmanouilidou et~al.}(2017)\citenamefont{Emmanouilidou,
  Shen, Deng, Chang, Shi, Kotliar, Xu, and
  Ni}}]{emmanouilidou2017magnetotransport}
\bibinfo{author}{\bibfnamefont{E.}~\bibnamefont{Emmanouilidou}},
  \bibinfo{author}{\bibfnamefont{B.}~\bibnamefont{Shen}},
  \bibinfo{author}{\bibfnamefont{X.}~\bibnamefont{Deng}},
  \bibinfo{author}{\bibfnamefont{T.-R.} \bibnamefont{Chang}},
  \bibinfo{author}{\bibfnamefont{A.}~\bibnamefont{Shi}},
  \bibinfo{author}{\bibfnamefont{G.}~\bibnamefont{Kotliar}},
  \bibinfo{author}{\bibfnamefont{S.-Y.} \bibnamefont{Xu}}, \bibnamefont{and}
  \bibinfo{author}{\bibfnamefont{N.}~\bibnamefont{Ni}},
  \bibinfo{journal}{Physical Review B} \textbf{\bibinfo{volume}{95}},
  \bibinfo{pages}{245113} (\bibinfo{year}{2017}).

\bibitem[{\citenamefont{Zhang et~al.}(2010)\citenamefont{Zhang, He, Chang,
  Song, Wang, Chen, Jia, Fang, Dai, Shan et~al.}}]{zhang2010crossover}
\bibinfo{author}{\bibfnamefont{Y.}~\bibnamefont{Zhang}},
  \bibinfo{author}{\bibfnamefont{K.}~\bibnamefont{He}},
  \bibinfo{author}{\bibfnamefont{C.-Z.} \bibnamefont{Chang}},
  \bibinfo{author}{\bibfnamefont{C.-L.} \bibnamefont{Song}},
  \bibinfo{author}{\bibfnamefont{L.-L.} \bibnamefont{Wang}},
  \bibinfo{author}{\bibfnamefont{X.}~\bibnamefont{Chen}},
  \bibinfo{author}{\bibfnamefont{J.-F.} \bibnamefont{Jia}},
  \bibinfo{author}{\bibfnamefont{Z.}~\bibnamefont{Fang}},
  \bibinfo{author}{\bibfnamefont{X.}~\bibnamefont{Dai}},
  \bibinfo{author}{\bibfnamefont{W.-Y.} \bibnamefont{Shan}},
  \bibnamefont{et~al.}, \bibinfo{journal}{Nature Physics}
  \textbf{\bibinfo{volume}{6}}, \bibinfo{pages}{584} (\bibinfo{year}{2010}).

\bibitem[{\citenamefont{Perdew et~al.}(1996)\citenamefont{Perdew, Burke, and
  Ernzerhof}}]{perdew1996generalized}
\bibinfo{author}{\bibfnamefont{J.~P.} \bibnamefont{Perdew}},
  \bibinfo{author}{\bibfnamefont{K.}~\bibnamefont{Burke}}, \bibnamefont{and}
  \bibinfo{author}{\bibfnamefont{M.}~\bibnamefont{Ernzerhof}},
  \bibinfo{journal}{Physical review letters} \textbf{\bibinfo{volume}{77}},
  \bibinfo{pages}{3865} (\bibinfo{year}{1996}).

\bibitem[{\citenamefont{Kresse and
  Furthm{\"u}ller}(1996{\natexlab{a}})}]{kresse1996efficiency}
\bibinfo{author}{\bibfnamefont{G.}~\bibnamefont{Kresse}} \bibnamefont{and}
  \bibinfo{author}{\bibfnamefont{J.}~\bibnamefont{Furthm{\"u}ller}},
  \bibinfo{journal}{Computational materials science}
  \textbf{\bibinfo{volume}{6}}, \bibinfo{pages}{15}
  (\bibinfo{year}{1996}{\natexlab{a}}).

\bibitem[{\citenamefont{Kresse and
  Furthm{\"u}ller}(1996{\natexlab{b}})}]{kresse1996efficient}
\bibinfo{author}{\bibfnamefont{G.}~\bibnamefont{Kresse}} \bibnamefont{and}
  \bibinfo{author}{\bibfnamefont{J.}~\bibnamefont{Furthm{\"u}ller}},
  \bibinfo{journal}{Physical review B} \textbf{\bibinfo{volume}{54}},
  \bibinfo{pages}{11169} (\bibinfo{year}{1996}{\natexlab{b}}).

\bibitem[{\citenamefont{Dahlmann and Schnering}(1973)}]{dahlmann1973cap}
\bibinfo{author}{\bibfnamefont{W.}~\bibnamefont{Dahlmann}} \bibnamefont{and}
  \bibinfo{author}{\bibfnamefont{H.}~\bibnamefont{Schnering}},
  \bibinfo{journal}{Naturwissenschaften} \textbf{\bibinfo{volume}{60}},
  \bibinfo{pages}{518} (\bibinfo{year}{1973}).

\bibitem[{\citenamefont{Rourke and Julian}(2012)}]{julian2012numerical}
\bibinfo{author}{\bibfnamefont{P.}~\bibnamefont{Rourke}} \bibnamefont{and}
  \bibinfo{author}{\bibfnamefont{S.}~\bibnamefont{Julian}},
  \bibinfo{journal}{Computer Physics Communications}
  \textbf{\bibinfo{volume}{183}}, \bibinfo{pages}{324} (\bibinfo{year}{2012}).

\bibitem[{\citenamefont{Giraldo-Gallo et~al.}(2016)\citenamefont{Giraldo-Gallo,
  Sangiorgio, Walmsley, Silverstein, Fechner, Riggs, Geballe, Spaldin, and
  Fisher}}]{giraldo2016fermi}
\bibinfo{author}{\bibfnamefont{P.}~\bibnamefont{Giraldo-Gallo}},
  \bibinfo{author}{\bibfnamefont{B.}~\bibnamefont{Sangiorgio}},
  \bibinfo{author}{\bibfnamefont{P.}~\bibnamefont{Walmsley}},
  \bibinfo{author}{\bibfnamefont{H.}~\bibnamefont{Silverstein}},
  \bibinfo{author}{\bibfnamefont{M.}~\bibnamefont{Fechner}},
  \bibinfo{author}{\bibfnamefont{S.}~\bibnamefont{Riggs}},
  \bibinfo{author}{\bibfnamefont{T.}~\bibnamefont{Geballe}},
  \bibinfo{author}{\bibfnamefont{N.}~\bibnamefont{Spaldin}}, \bibnamefont{and}
  \bibinfo{author}{\bibfnamefont{I.}~\bibnamefont{Fisher}},
  \bibinfo{journal}{Physical Review B} \textbf{\bibinfo{volume}{94}},
  \bibinfo{pages}{195141} (\bibinfo{year}{2016}).

\bibitem[{\citenamefont{Mostofi et~al.}(2008)\citenamefont{Mostofi, Yates, Lee,
  Souza, Vanderbilt, and Marzari}}]{mostofi2008wannier90}
\bibinfo{author}{\bibfnamefont{A.~A.} \bibnamefont{Mostofi}},
  \bibinfo{author}{\bibfnamefont{J.~R.} \bibnamefont{Yates}},
  \bibinfo{author}{\bibfnamefont{Y.-S.} \bibnamefont{Lee}},
  \bibinfo{author}{\bibfnamefont{I.}~\bibnamefont{Souza}},
  \bibinfo{author}{\bibfnamefont{D.}~\bibnamefont{Vanderbilt}},
  \bibnamefont{and} \bibinfo{author}{\bibfnamefont{N.}~\bibnamefont{Marzari}},
  \bibinfo{journal}{Computer physics communications}
  \textbf{\bibinfo{volume}{178}}, \bibinfo{pages}{685} (\bibinfo{year}{2008}).

\bibitem[{\citenamefont{Wu et~al.}(2017)\citenamefont{Wu, Zhang, Song, Troyer,
  and Soluyanov}}]{wu2017wanniertools}
\bibinfo{author}{\bibfnamefont{Q.}~\bibnamefont{Wu}},
  \bibinfo{author}{\bibfnamefont{S.}~\bibnamefont{Zhang}},
  \bibinfo{author}{\bibfnamefont{H.-F.} \bibnamefont{Song}},
  \bibinfo{author}{\bibfnamefont{M.}~\bibnamefont{Troyer}}, \bibnamefont{and}
  \bibinfo{author}{\bibfnamefont{A.~A.} \bibnamefont{Soluyanov}},
  \bibinfo{journal}{arXiv preprint arXiv:1703.07789}  (\bibinfo{year}{2017}).

\bibitem[{\citenamefont{Narayan et~al.}(2014)\citenamefont{Narayan, Di~Sante,
  Picozzi, and Sanvito}}]{narayan2014topological}
\bibinfo{author}{\bibfnamefont{A.}~\bibnamefont{Narayan}},
  \bibinfo{author}{\bibfnamefont{D.}~\bibnamefont{Di~Sante}},
  \bibinfo{author}{\bibfnamefont{S.}~\bibnamefont{Picozzi}}, \bibnamefont{and}
  \bibinfo{author}{\bibfnamefont{S.}~\bibnamefont{Sanvito}},
  \bibinfo{journal}{Physical review letters} \textbf{\bibinfo{volume}{113}},
  \bibinfo{pages}{256403} (\bibinfo{year}{2014}).

\end{thebibliography}

\end{document}